\documentclass[RNAAS]{aastex631}

\usepackage{comment}
\shorttitle{Validating Gaia DR3 Candidate Variables with TESS}
\shortauthors{Zhou A.-Y.}
\graphicspath{{./}{figures/}}
\begin{document}

\title{Validating Gaia DR3 Pulsating Variable Classifications with \textit{TESS}: 
Building Reliable $\delta$ Scuti and $\gamma$ Doradus Stars Catalogs (In Progress)\footnote{Dedicated to my wife Jingyun Zhang}}

\correspondingauthor{Ai-Ying Zhou}	
\email{aiying@nao.cas.cn}

\author[0000-0002-2919-6989]{Ai-Ying Zhou}
\affiliation{National Astronomical Observatories, Chinese Academy of Sciences, \\
A20 Datun Road, Chaoyang District, Beijing 100101, P.R. China}
\affiliation{Key Laboratory of Radio Astronomy and Technology (CAS)}

\begin{abstract}
Gaia DR3 revealed 748,058 pulsating variable stars of mixed ${\rm DSCT|GDOR|SXPHE}$ types.  
This project undertakes a comprehensive examination to validate and distinguish these stars using \textit{TESS} data. 
Aiming for reliable catalogs of bona fide $\delta$ Scuti and $\gamma$ Doradus stars, I have validated 
1715 $\delta$ Scuti stars,
1403 $\gamma$ Doradus stars,
and identified 260 eclipsing binaries, 
one RR Lyrae star, and 460+ rotating variables from an initial sample of 16,690 objects. 
Notably,  15 of the newfound eclipsing binaries 
harbor pulsating $\gamma$ Doradus components.

\end{abstract}
%% Limited word for Abstract: 150 words
%% Remember, keep the abstract concise and informative, highlighting the project's objectives, methods, and potential significance.

%% The AAS Journals now uses Unified Astronomy Thesaurus concepts:  https://astrothesaurus.org
%% but this old "keyword" functionality is maintained in case authors want to include these concepts in their preprints.
%--- RR Lyrae variable stars(1410)
%--- Ellipsoidal variable stars(455)
%--- Hertzsprung Russell diagram(725)	
%--- CCD photometry(208)	
%--- Stellar evolution(1599)	Stellar structures(1631)		
%--- Stellar oscillations(1617)	--- Asteroseismology(73)
%% https://astrothesaurus.org/concept-select/

\section{Introduction} 
\label{sec:intro}
\textit{Gaia} Data Release 3 %Part 4 Variability
released 12.4 million variables,  
including 748,058 pulsating stars of mixed ${\rm DSCT|GDOR|SXPHE}$ types among the variability classification results of all classifiers -- 9,976,881 objects~\citep[][]{GaiaDR3}
%  (in the file {\tt vclassre.dat}
\footnote{https://cdsarc.cds.unistra.fr/viz-bin/cat/I/358}.

With its extensive catalog of variable stars, \textit{Gaia} DR3   
presents a transformative opportunity 
to explore the properties of $\delta$ Scuti and $\gamma$ Doradus stars.
However, independent verification using dedicated observations remains crucial, as the catalog includes known variables, eclipsing binaries, and non-variables (evident from a trial screening of the leading 16,690 objects,  %a sample of about 
$\sim$2.2\% of total).

NASA's Transiting Exoplanet Survey Satellite (\textit{TESS},~\citealt{TESS}), 
provides exquisite photometry, 
making it ideal to identify stellar variability.  
This project leverages the synergy of \textit{Gaia} catalogs and \textit{TESS} high-precision light curves for a rigorous verification campaign of these `{\tt Gaia variables}'. 
My approach combines visual inspection of light curves, periodograms, and stellar parameters evaluation for stars location on the Hertzsprung-Russell diagram 
to validate variability and identify genuine pulsators. 
This process improves \textit{Gaia} variable catalogs by removing false positives, 
yielding reliable catalogs of confirmed DSCT and GDOR.

\section{Sample and Data}
This project works on two sources:
\begin{itemize}
\item 748,058 pulsating variable stars of mixed `${\rm DSCT|GDOR|SXPHE}$' types from  
\textit{Gaia} DR3
% Part 4 Variability classification results of all classifiers (9,976,881 objects) 
\citep{GaiaDR3}; 
\item 8,882 `DSCT\_SXPHE'  and 3,018 stars of short time-scale from the \textit{Gaia} DR2 Variability Results of 363,969 records \citep{GaiaDR2}. 
\end{itemize}
Candidates were chosen according to the following selection criteria: 
effective temperature 5,500 to 12,000\,K, 
brighter than \textit{TESS} magnitude 16\fm0, 
and in 16\fm0--19\fm0 with `score of the best class' higher than 0.3.

For each star, I downloaded the shortest-exposure light curve from the latest observations (Sectors 1-71), prioritizing SPOC, TESS-SPOC, QLP, and TASOC products available at Mikulski Archive for Space Telescopes (MAST\footnote{https://mast.stsci.edu/portal/Mashup/Clients/Mast/Portal.html}):~  
\begin{itemize}
\item 20-second 
%cadence observations 
%(\href{https://archive.stsci.edu/tess/bulk_downloads/bulk_downloads_ffi-tp-lc-dv.html#fastlc}{DOI:10.17909/t9-st5g-3177})
and 2-minute cadence observations
% (\href{https://archive.stsci.edu/tess/bulk_downloads/bulk_downloads_ffi-tp-lc-dv.html#lc}{DOI:10.17909/t9-nmc8-f686})
~\citep{https://doi.org/10.17909/t9-st5g-3177, https://doi.org/10.17909/t9-nmc8-f686}, 
\item MIT Quick‑Look Pipeline% (DOI: 10.17909/t9-r086-e880)
~\citep{https://doi.org/10.17909/t9-r086-e880}, 
\item \textit{TESS} Light Curves From Full Frame Images% (TESS‑SPOC, DOI: 10.17909/t9-wpz1-8s54)
~\citep[TESS-SPOC,][]{https://doi.org/10.17909/t9-wpz1-8s54}, 
\item \textit{TESS} Data For Asteroseismology Lightcurves% (TASOC, DOI: 10.17909/t9-4smn-dx89)
~\citep[TASOC, ][]{https://doi.org/10.17909/t9-4smn-dx89}. 
\end{itemize}
Astronomical parameters are based on Simbad, 
 \textit{TESS} Input Catalog ~\citep[TIC v8.2,][]{TICv82} and 
\textit{Gaia} DR2/DR3~\citep[][]{GaiaDR3, GaiaMission}.
Spectral types were empirically derived using effective temperatures if unavailable.

%% obtain bibTeX entries for a DOI: 
%% curl -k -LH "Accept: application/x-bibtex" https://doi.org/10.17909/t9-r086-e880

\section{Methodology and Identification} 
\label{sec:data}

A star's variability type is determined on following factors:
morphological characteristics and periodograms of their light curves,   
astrophysical parameters and their H-R diagram location.

I first check each candidate against the most relative catalogues:  
 70\,552 DSCT and  8\,080 GDOR~\citep{zhou2024};
123\,841 and 84\,206 {\it TESS} variables~\citep{2022arXiv221210776B, 2023ApJS..268....4F}; 
and the new variables identified  by \citet{zhou2023c,zhou2023d,zhou2023g}. 
Then I check %the variability status of each candidate 
against Simbad.
Candidates already identified as known variables in existing catalogs 
%including those labeled `PulsV*', `Variables*', `EclBin', etc by Simbad, 
and those lacking corresponding \textit{TESS} data are excluded.
Each candidate underwent a two-step confirmation process: 
visually inspect the downloaded light curves, 
%one of the \textit{TESS} sectors  with the shortest cadence available  at MAST; 
retrieve stellar parameters 
%(effective temperature, luminosity, surface gravity, mass, etc.) 
and evaluate their locations on H-R diagram.

A custom Python program (adapted from \citealt{zhou2023c}) facilitated the interactive screening, 
which performed various tasks: 
Online data retrieval from Simbad, TIC, MAST, and \textit{Gaia};
Data processing and visualization;
% of SAP and PDCSAP, 10-minute and 30-minute binned light curves; (detrending, outliers removal, flattening, binning);
Periodogram and noise-level computation; 
Significant peaks identification; % (over a signal-to-noise ratio of 4.5);
Cross-matching \textit{Gaia} DR3 identifiers with TIC and Simbad main identifiers. % (within 3 arcseconds).

\section{Results}
\label{results}

An initial cross-match of 16,690 pulsating stars from \textit{Gaia} DR3 against Simbad and known catalogs revealed:
1454 pulsating variables (`PulsV', 8.7\%, e.g. NP Del,  HD 83628); 
%, TYC 4189-610-1, HD 40800, HD 51832, HD 80366,   HD 109838, TYC 4110-508-1, 
143  eclipsing binaries (`EclBin', 0.86\%, e.g. FI Boo, V452 Cam); %, V1208 Her, TYC 4529-785-1, 
346 generic variables (`Variable', 2.1\%); %, e.g. ATO J259.2541+46.6265 ); 
1108 known $\delta$ Sct (6.6\%, e.g. GG UMa, V1209 Her); %LV UMa, CQ Lyn, 
68 known $\gamma$ Dor (0.41\%, e.g. LS UMa, LU UMa). %, KO UMa. 
However, some Simbad `PulsV' classifications potentially relied on \textit{Gaia} DR3. 
%results and uses \textit{Gaia} DR3 source ID as main identifiers (e.g. Gaia DR3 1048928595340879616, Gaia DR3 1005246651932504448, etc.).

\begin{figure*}
   \vspace{-2mm}
   \centering
   \includegraphics[width=1.06\textwidth, height=239mm,angle=0,scale=1.00]{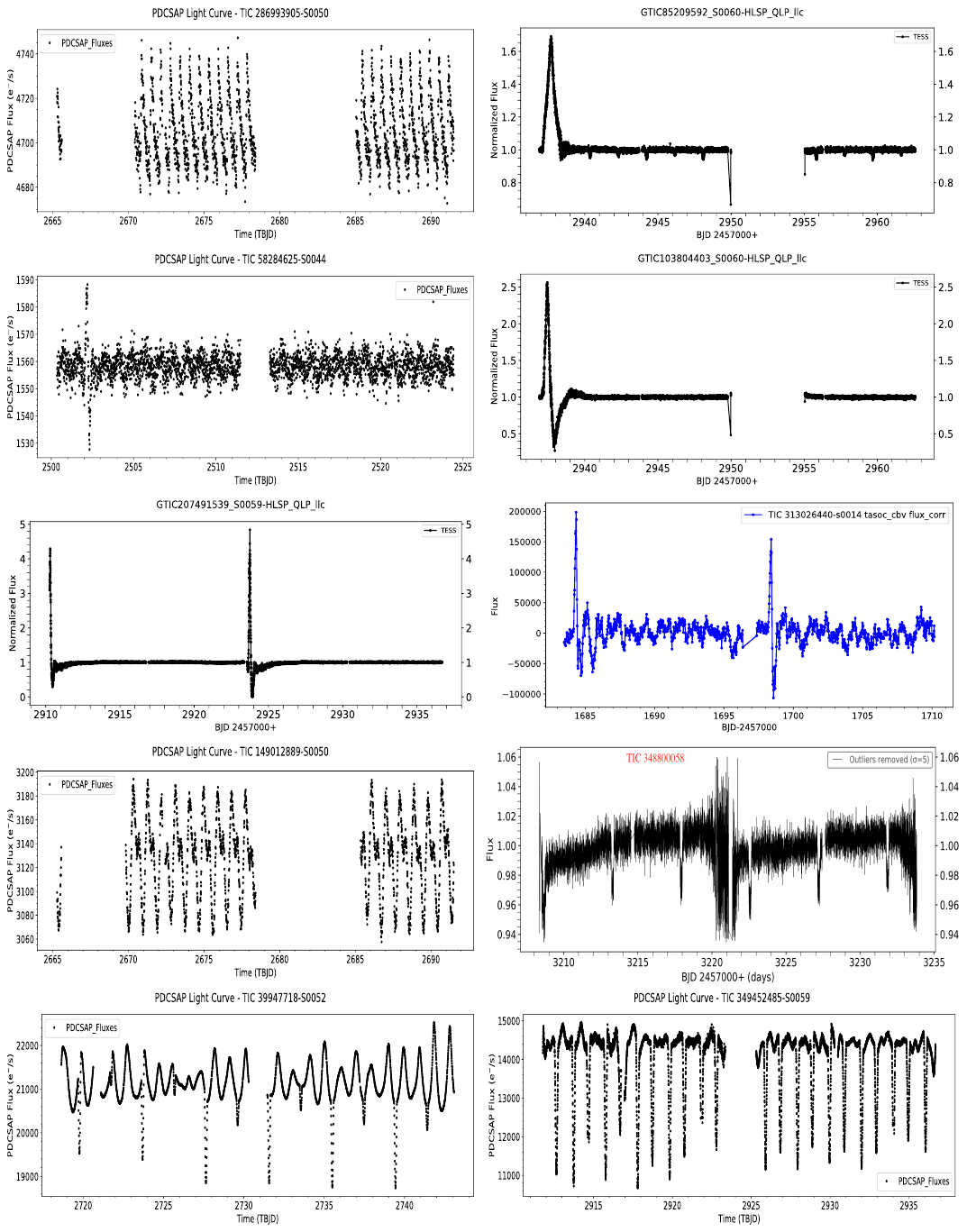}
\begin{minipage}[]{0.950\textwidth}
   \caption{Example \textit{TESS} light curves of ten newly identified variables 
   not ${\rm DSCT|GDOR|SXPHE}$ types. 
Top left: a new RRab star; Top right: EA+flare;
2nd and 3rd rows: four new heartbeat stars; 
4th row left: a new rotating ellipsoidal variable star; Right, a candidate planetary transit;
5th rows: two new eclipsing binaries with pulsating $\gamma$ Dor primary components.}
   \label{fig:GDR3-TESS}
\end{minipage}
\end{figure*}

\textit{TESS} light curves screened 7,360 objects out of 16,690 (remaining lacked \textit{TESS} data or were not suitable), validated  following \textit{Gaia} DR3 discoveries:  
1715 $\delta$ Sct, 
1403 $\gamma$ Dor, 
one RR Lyr star (=TYC 2553-889-1),
460+ rotating variables, 
260+ eclipsing binaries (including 15 with pulsating GDOR components), and others.  
Additionally, 
760+ known variables, 
2600+ unclassified variables, 
and 1050+ non-variables were identified at the precision of \textit{TESS} photometry.
Figure~\ref{fig:GDR3-TESS} shows ten of the \textit{TESS}-validated `{\tt \textit{Gaia} variables}' that actually do not belong to ${\rm DSCT|GDOR|SXPHE}$ types. 
Current validated variables as ``Version 1" is available 
at Zenodo \href{https://zenodo.org/doi/10.5281/zenodo.10023631}{DOI:10.5281/zenodo.10023631} and \href{https://deltascuti.wixsite.com/delta/download}{this website} too.

\section{Discussion}
The preliminary results of initial examination of 16,690 objects from 
\textit{Gaia} DR3 pulsating variables of ${\rm DSCT|GDOR|SXPHE}$ types using \textit{TESS} data, 
emphasizes the importance of cross-validation 
with complementary data sources for robust scientific results.
While automated algorithms are essential for classifying massive datasets, critical evaluation is crucial to avoid limitations.

Assuming current validation rates hold for the remaining sample, 
I would expect to validate over 
85,000 $\delta$ Scuti and 66,000 $\gamma$ Doradus stars,   
exceeding a success rate of 20\% for the whole sample. 
The number of stars that can be identified and confirmed ultimately depends on \textit{TESS} data coverage. 
This is an ongoing work, I shall provide updates as the validation progresses.

%============================================== Figure 

\begin{acknowledgments}
I am indebted to my wife Jingyun Zhang for her unwavering support throughout my research.
This work includes data collected with \textit{TESS} mission,
I acknowledge the use of \textit{TESS} data, which are derived from
pipelines at \textit{TESS} Science Processing Operations Center.
 \textit{TESS} High Level Science Products produced by the Quick-Look Pipeline at
 \textit{TESS} Science Office at MIT, which are publicly available from the Mikulski Archive for Space Telescopes data archive at the Space Telescope Science Institute.
Funding for \textit{TESS} mission is provided by NASA Explorer Program.
STScI is operated by the Association of Universities for Research in Astronomy, Inc., under NASA contract NAS 5–26555.
This research has made use of
SIMBAD, operated at CDS, Strasbourg, France; 
data from the European Space Agency mission {\it Gaia}, processed by the {\it Gaia}
Data Processing and Analysis Consortium.
Funding for the DPAC has been provided by national institutions, in particular the institutions
participating in the {\it Gaia} Multilateral Agreement.
\software{Astropy \citep{astropy:2013, astropy:2018, astropy:2022}; 
Astroquery~\citep{astroquery:2019};
Lightkurve \citep{lightkurve}.}
\end{acknowledgments}
%This research made use of Lightkurve, a Python package for Kepler and TESS data analysis (Lightkurve Collaboration, 2018).
%This work made use of Astropy:\footnote{http://www.astropy.org} a community-developed core Python package and an ecosystem of tools and resources for astronomy \citep{astropy:2013, astropy:2018, astropy:2022}.

\begin{comment}
\end{comment}

\bibliography{../raa/zayRef}{}
\bibliographystyle{aasjournal}

\end{document}